\documentclass{elsart}
\usepackage{epsfig}

\newcommand{\eff}{\varepsilon}

\newcommand{\jpsi}{J/\psi}

\newcommand{\pip}{\pi^+}
\newcommand{\pin}{\pi^-}

\newcommand{\beq}{\begin{equation}}
\newcommand{\eeq}{\end{equation}}
\newcommand{\beqn}{\begin{eqnarray}}
\newcommand{\eeqn}{\end{eqnarray}}
\newcommand{\beqns}{\begin{eqnarray*}}
\newcommand{\eeqns}{\end{eqnarray*}}
\newcommand{\bfg}{\begin{figure}}
\newcommand{\efg}{\end{figure}}
\newcommand{\bitm}{\begin{itemize}}
\newcommand{\eitm}{\end{itemize}}
\newcommand{\bnum}{\begin{enumerate}}
\newcommand{\enum}{\end{enumerate}}
\newcommand{\btbl}{\begin{table}}
\newcommand{\etbl}{\end{table}}
\newcommand{\btbu}{\begin{tabular}}
\newcommand{\etbu}{\end{tabular}}

\newcommand{\kp}{K^+}
\newcommand{\kn}{K^-}
\newcommand{\ks}{K^{0}_{S}}

\newcommand{\etap}{\eta^{\prime}}
\newcommand{\rhoo}{\rho^0}
\newcommand{\g}{\gamma}
\newcommand{\be}{\begin{enumerate}}
\newcommand{\ee}{\end{enumerate}}
\newcommand{\bi}{\begin{itemize}}
\newcommand{\ei}{\end{itemize}}

\newcommand{\rar}{\rightarrow}

\newcommand{\pio}{\pi^0}

\newcommand{\ar}{\rightarrow}

\renewcommand{\thefootnote}{\fnsymbol{footnote}}

\begin{document}

\begin{frontmatter}
\title{\boldmath  Measurements of  $\jpsi$ Decays into
$2(\pip\pin)\eta$ and $3(\pip\pin)\eta$
}

\date{15 Jan. 2004}

\maketitle

\begin{center}
M.~Ablikim$^{1}$,    J.~Z.~Bai$^{1}$,               Y.~Ban$^{11}$,
J.~G.~Bian$^{1}$,    X.~Cai$^{1}$,                  J.~F.~Chang$^{1}$,
H.~F.~Chen$^{17}$,   H.~S.~Chen$^{1}$,              H.~X.~Chen$^{1}$,
J.~C.~Chen$^{1}$
% ...
\\(BES Collaboration)\\
\vspace{0.2cm}
$^{1}${\it Institute of High Energy Physics, Beijing 100049, People's
Republic of
 China }\\
% ...
$^{11}$ {\it Peking University, Beijing 100871, People's Republic of
China}\\
% ...
$^{17}$ {\it University of Science and Technology of China, Hefei 230026,
People's Republic of
China}\\
% ...

\end{center}
\vskip 0.3cm


\begin{abstract}
Based on a sample of $5.8\times 10^7$ $\jpsi$ events taken with the
BESII detector, the branching fractions of $\jpsi\to 2(\pip\pin)\eta$
and $\jpsi\to 3(\pip\pin)\eta$ are measured for the first time to be
$(2.26\pm0.08\pm 0.27)\times 10^{-3}$ and $(7.24\pm0.96\pm1.11)\times
10^{-4}$, respectively.
\vspace{3\parskip}


\end{abstract}
\end{frontmatter}
\clearpage

\section{Introduction}   \label{introd}
More than one hundred exclusive decay modes of the 
$\jpsi$
have been reported  since its discovery
at Brookhaven~\cite{aube} and SLAC~\cite{augu} in 1974. According to
Ref.~\cite{physrep}, direct
hadronic, electromagnetic and radiative decays represent roughly
65\%, 14\%, and 7\% of the total $\jpsi$ decay rate, respectively. 
Up to now, only about half of all hadronic decays, 34.8\%, have been
measured in exclusive reactions.
The sample of 58 million $\jpsi$ events, taken at BESII, provides a chance to
measure some of the missing hadronic decays.
In this analysis, we report  the first measurements of
$\jpsi\ar 2(\pip\pin)\eta$ and $\jpsi\ar 3(\pip\pin)\eta$.



The upgraded Beijing Spectrometer (BESII) detector located
at the Beijing Electron-Positron Collider (BEPC)
is a large solid-angle
magnetic spectrometer which is described in detail in Ref.~\cite{besii}.
The momentum of the charged particle is determined by a
40-layer cylindrical main drift chamber (MDC) which has a momentum
resolution of
 $\sigma_{p}$/p=$1.78\%\sqrt{1+p^2}$ ($p$ in GeV/c).
Particle identification is accomplished by specific ionization ($dE/dx$)
measurements in the drift chamber and time-of-flight (TOF) information in
a barrel-like array of 48 scintillation counters. The $dE/dx$ resolution
is $\sigma_{dE/dx}=8.0\%$; the TOF resolution for Bhabha events is
$\sigma_{TOF}=180$ ps.
Radially outside of the time-of-flight counters is a 12-radiation-length
barrel shower counter (BSC) comprised of gas
tubes interleaved with lead sheets. The BSC measures
the energy and direction of photons with resolutions
of $\sigma_{E}/E\simeq21\%\sqrt{E}$ ($E$ in GeV), $\sigma_{\phi}=7.9$ mrad, and
$\sigma_{z}=2.3$ cm. The iron flux return of the magnet is instrumentd
with three double layers of counters that are used to identify muons.

A GEANT3 based Monte Carlo package (SIMBES) with detailed
consideration of the detector performance
is used. The consistency between data and Monte Carlo has been carefully
checked in many high purity physics channels, and the agreement is
reasonable.




\section{\boldmath Analysis of  $\jpsi\rar 2(\pip\pin)\eta$}
\label{4pieta}
This decay is observed in the topology $\pip\pin\pin\pin\g\g$.
Events with four charged tracks and at least two isolated photons
are selected. The  selection criteria for charged tracks and photons
are described in detail in Ref.~\cite{rhopi}. Each charged track must be
well fitted to a helix, originating from the interaction region of R$_{xy}$
$<$ 0.02 m and $|z|~<$ 0.2m, and have a polar angle, $\theta$, in the range
$|\cos\theta|$ $<$ 0.8. 

Isolated photons
are those that have energy deposited in the BSC greater than 60 MeV, the
angle between the direction at the first layer of the BSC and the developing
direction of the cluster less than 30$^{\circ}$, and the angle between
photons and any charged tracks larger than $10^{\circ}$.
 To eliminate tracks from $\gamma$ conversions,
the minimum angle between any two oppositely-charged tracks is
required to be
greater than $10^{\circ}$.

A 4C kinematic fit is  performed under the hypothesis
$\pip\pin\pip\pin\g\g$,
and the chi-squared, $\chi^{2}_{\pip\pin\pip\pin\g\g}$, is required to be less than 15. 
$\chi^{2}_{\pip\pin\pip\pin\g\g}$ is also required to be less than the
chi-squares for the
$\kp\kn\pip\pin\g\g$ and $\pip\pin\pip\pin\g\g\g$
hypotheses.
\begin{figure}[htpb]
\centerline{\includegraphics[width=0.5\textwidth,height=0.38\textheight]
        {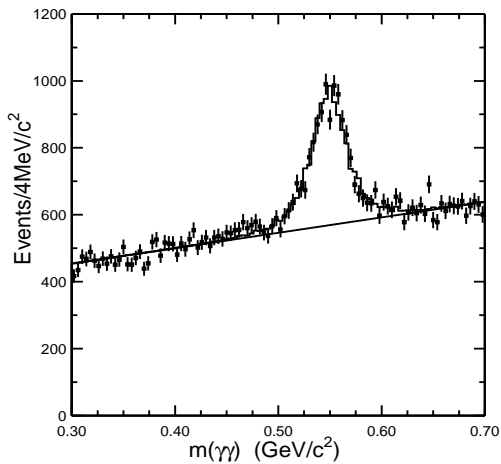}}
\caption{The distribution of $m_{\g\g}$ for candidate $J/\psi \ar
  \eta 2(\pi^+ \pi^-)$ events. Dots with error bars
are data, the histogram is the  shape of the $\eta$ from Monte Carlo
simulation, 
and the curve is background.}
\label{4pietafit}
\end{figure}

\begin{figure}[htpb]
\begin{center}
\includegraphics[width=0.7\textwidth,height=0.26 \textheight]
        {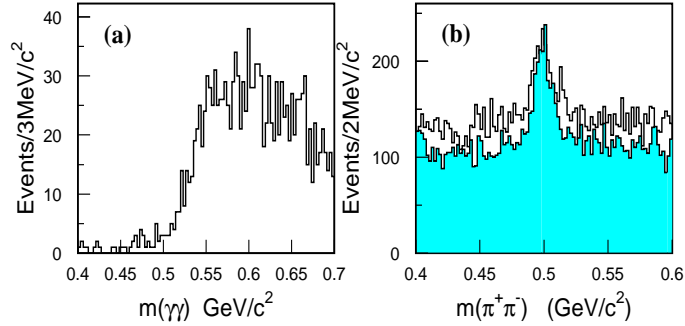}
\caption{(a) Distribution of $m_{\g\g}$ from Monte Carlo simulation of
$\jpsi\ar\g 2(\pip\pin)\eta$. (b) Distribution of $m_{\pip\pin}$ from
the $\eta$ signal region (blank histogram) and $\eta$ sidebands
(shaded histogram) for $J/\psi \ar \gamma\gamma 2(\pi^+ \pi^-)$
events.}
\label{metapipi}
\end{center}

\end{figure}

 After the above selection, the $m_{\g\g}$ distribution is shown in 
Fig.~\ref{4pietafit}, where a clear $\eta$ signal is observed.
Monte Carlo simulation is used to estimate the background,
and backgrounds from simulated channels are listed in Table~\ref{bkg},
where $N^{MC}_{sel}$ is the number of events after event selection and
$N^{norm}$ is the background normalized to 58 million $\jpsi$ events.
The sum of background events is 18 events, which can be ignored.
Another possible background channel is from $\jpsi\ar\g\eta 2(\pip\pin)$.
No obvious $\eta$ signal is seen in the $m_{\g\g}$ distribution from 
$\jpsi\ar\g\eta 2(\pip\pin)$, as shown
in
Fig.~\ref{metapipi}(a). Therefore the background from this channel 
can also be ignored.




\begin{table}[htpb]
\caption{Background estimates for the $J/\psi \ar \eta 2(\pi^+ \pi^-)$
  case. $N^{MC}_{sel}$ are the number of events
passing selection criteria; $N^{norm}$ is the background normalized to
58 million  $J/\psi$ events.}
\begin{center}
\begin{tabular}{c|c|c|c}
\hline
Channel & MC sample &$N^{MC}_{sel}$ & $N^{norm}$\\
\hline
$\jpsi\ar\phi\eta$ & 40,000 &0 & 0\\
$\jpsi\ar\rho\eta$ & 40,000 & 0 & 0\\
$\jpsi\ar\g\eta\pip\pin$ & 70,000 & 2 & 3\\
$\jpsi\ar\omega\eta(\eta\ar\g\g)$ & 40,000 & 11 & 8\\
$\jpsi\ar\omega\etap(\etap\ar\pip\pin\eta)$ &100,000 & 135& 2\\
$\jpsi\ar\phi\etap(\etap\ar\pip\pin\eta)$ & 40,000 & 2 & 0\\
$\jpsi\ar\g\etap(\etap\ar\pip\pin\eta)$ & 50,000 & 6 & 5 \\
\hline
\end{tabular}
\end{center}
\label{bkg}
\end{table}

Background from events with a $\ks$ in the final state are estimated
 from $\eta$ sidebands.  Fig.~\ref{metapipi}(b) shows the mass
 distribution of all $\pip\pin$ pairs for events with
 $m_{\gamma\gamma}$ in the $\eta$ region
 ($|m_{\g\g}-0.55|<0.05$~GeV/$c^2$), and the shaded histogram is for
 $\eta$ sidebands ($0.45$~GeV/$c^2$ $<m_{\g\g}<0.50$ GeV/$c^2$ and
 0.60 GeV/$c^2$ $<m_{\g\g}<$0.65 GeV/$c^2$).  From
 Fig.~\ref{metapipi}, we conclude that the $\ks$ signals are consistent
with coming from background, as estimated 
from $\eta$ sideband events. 

The  $m_{\g\g}$ distribution, shown in Fig.~\ref{4pietafit}, is fitted
with a Monte Carlo determined  shape for the $\eta$ and a second
order polynomial and yields
4839$\pm$158  $J/\psi \ar 2(\pi^+ \pi^-) \eta$, $\eta \ar \gamma
\gamma$ events.

\section{\boldmath Analysis of $\jpsi\ar 3(\pip\pin)\eta$}
\label{6pieta}
Events with six good charged tracks and at least two isolated photons
 are selected. The angle between two oppositely charged tracks is required to
 be greater than $10^{\circ}$ to remove $\gamma$ conversions.  A four
 constraint kinematic fit is made to the hypothesis $J/\psi \ar
 3(\pip\pin)\g\g$, and $\chi^{2}_{\pip\pin\pip\pin\pip\pin\g\g}$ is
 required to be less than 15.

\begin{figure}[htpb]
\centerline{\includegraphics[width=0.5\textwidth,height=0.38\textheight]
        {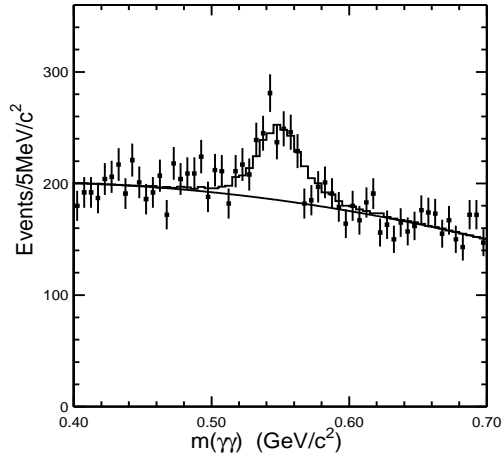}}
\caption{Distribution of $m_{\g\g}$ for  $J/\psi \ar
 3(\pip\pin)\g\g$ candidate events. Dots with error bars
are data, the histogram is the $\eta$ shape determined by Monte Carlo
 simulation, 
and the curve is background.}
\label{eta7pi}
\end{figure}

After the above selection, the two photon invariant mass distribution
is shown in Fig. \ref{eta7pi}; an $\eta$ signal is evident.



\begin{figure}[htpb]

\begin{center}
\includegraphics[width=0.7\textwidth,height=0.26\textheight]
        {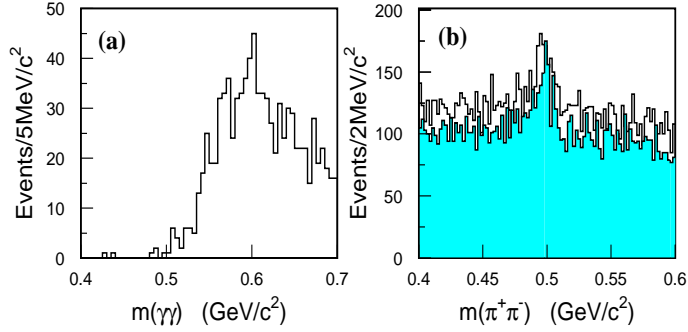}
\caption{(a) The $m_{\g\g}$ distribution from Monte Carlo simulation
of $\jpsi\ar\g 3(\pip\pin)\eta$; (b) The distribution of
$m_{\pip\pin}$ from $\eta$ signal events (histogram) and $\eta$
sidebands (shaded histogram) for $J/\psi \ar \gamma \gamma 3(\pi^+
\pi^-)$ events.}
\label{m6etapipi}
\end{center}

\end{figure}

Monte Carlo simulation indicates that background from the decay modes
 listed in Table~\ref{bkg} can be ignored.  Other possible backgrounds
 are from $\jpsi\ar\g 2(\pip\pin)\eta$ and $\jpsi\ar\g
 3(\pip\pin)\eta$ events. As described in Section~\ref{4pieta}, the
 $m_{\g\g}$ distribution from $\jpsi\ar\g 2(\pip\pin)\eta$ shows no
 clear $\eta$ peak.
For $\jpsi\ar\g 3(\pip\pin)\eta$,
the  $m_{\g\g}$ distribution from Monte Carlo simulation, shown in  
Fig.~\ref{m6etapipi}(a), 
also does not show a peak in the $\eta$ region, so its contribution can
also be ignored.

The background with $\ks$ final states can also be estimated, as was
done previously, using the $\eta$
sidebands. Fig.~\ref{m6etapipi}(b) shows the $\pip\pin$ mass
distribution. The full histogram is the $m_{\pip\pin}$ distribution
for events in the $\eta$ signal region ($|m_{\g\g}-0.55|<$0.05
GeV/$c^2$), and the shaded histogram is for events from the $\eta$
sidebands (0.45 GeV/$c^2 <m_{\g\g}<$ 0.50 GeV/$c^2$ and 0.60
GeV$/c^2<m_{\g\g}<$ 0.65 GeV/$c^2$).  As in Section~\ref{4pieta}, we
can conclude from Fig.~\ref{m6etapipi}(b) that most events with $\ks$
are associated with backgrounds which are measured by $\eta$
sidebands.

Fitting the $\g\g$ mass distribution in Fig.~\ref{eta7pi}  with the
Monte Carlo shape for $\eta$
and a second order polynomial, as described in Section~\ref{4pieta}, yields
616$\pm$ 82 events.

\section{Detection efficiency}
\begin{figure}[htpb]
\begin{minipage}[t]{6.5cm}
\begin{center}

\centerline{\includegraphics[width=0.8\textwidth,height=0.3\textheight]
        {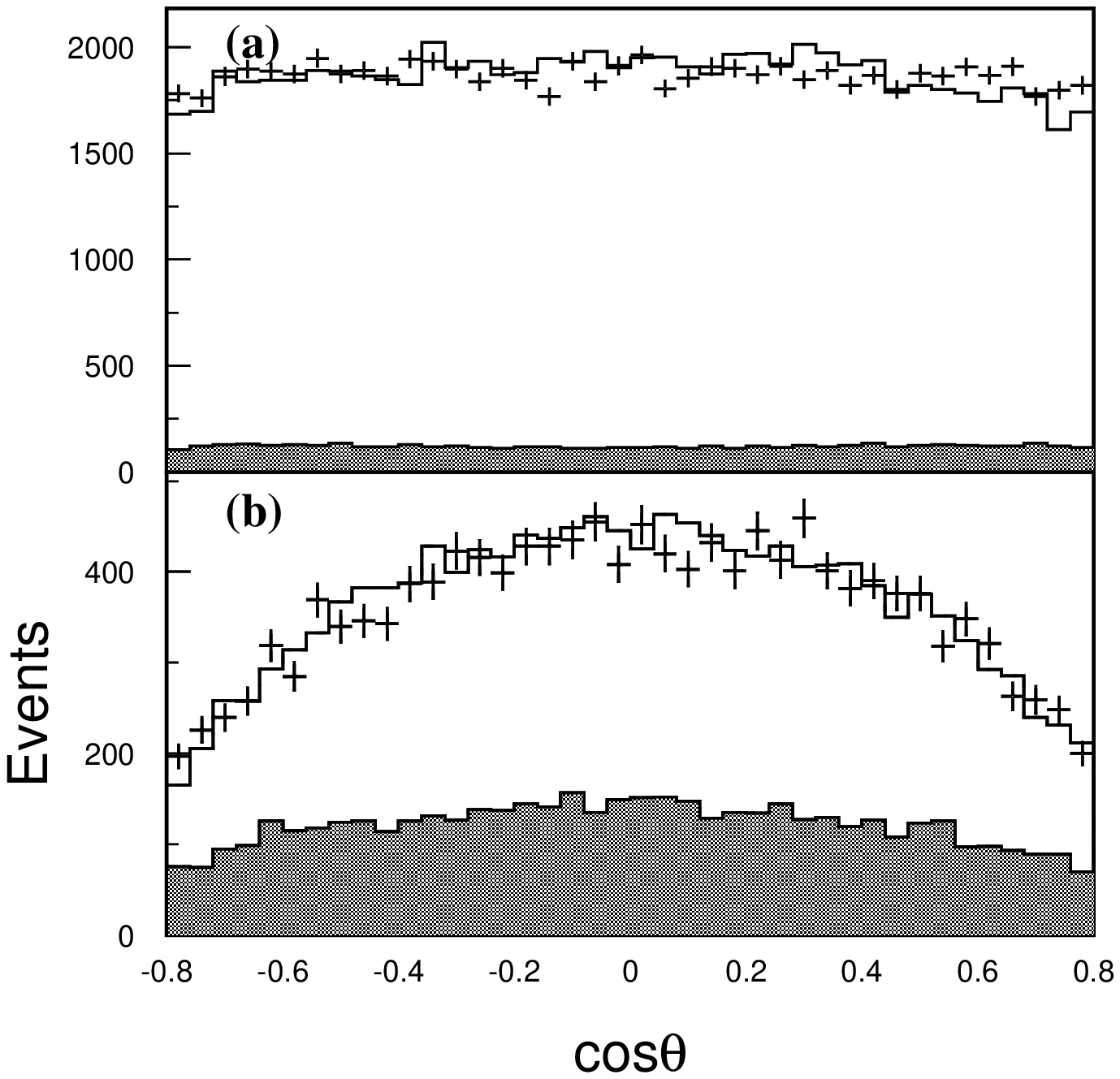} }
\caption{The $\cos\theta$ distribution of (a) charged tracks and (b)
$\eta$ particles, where the crosses are data, the shaded histogram is from Monte
Carlo simulation of $\jpsi\ar 2(\pip\pin)\eta$, and the full histogram
is the sum of sideband background and Monte Carlo simulation.}
\label{4picos}
\end{center}
\end{minipage}
\hfill
\begin{minipage}[t]{6.5cm}
\begin{center}
\includegraphics[width=0.8\textwidth,height=0.3\textheight]
        {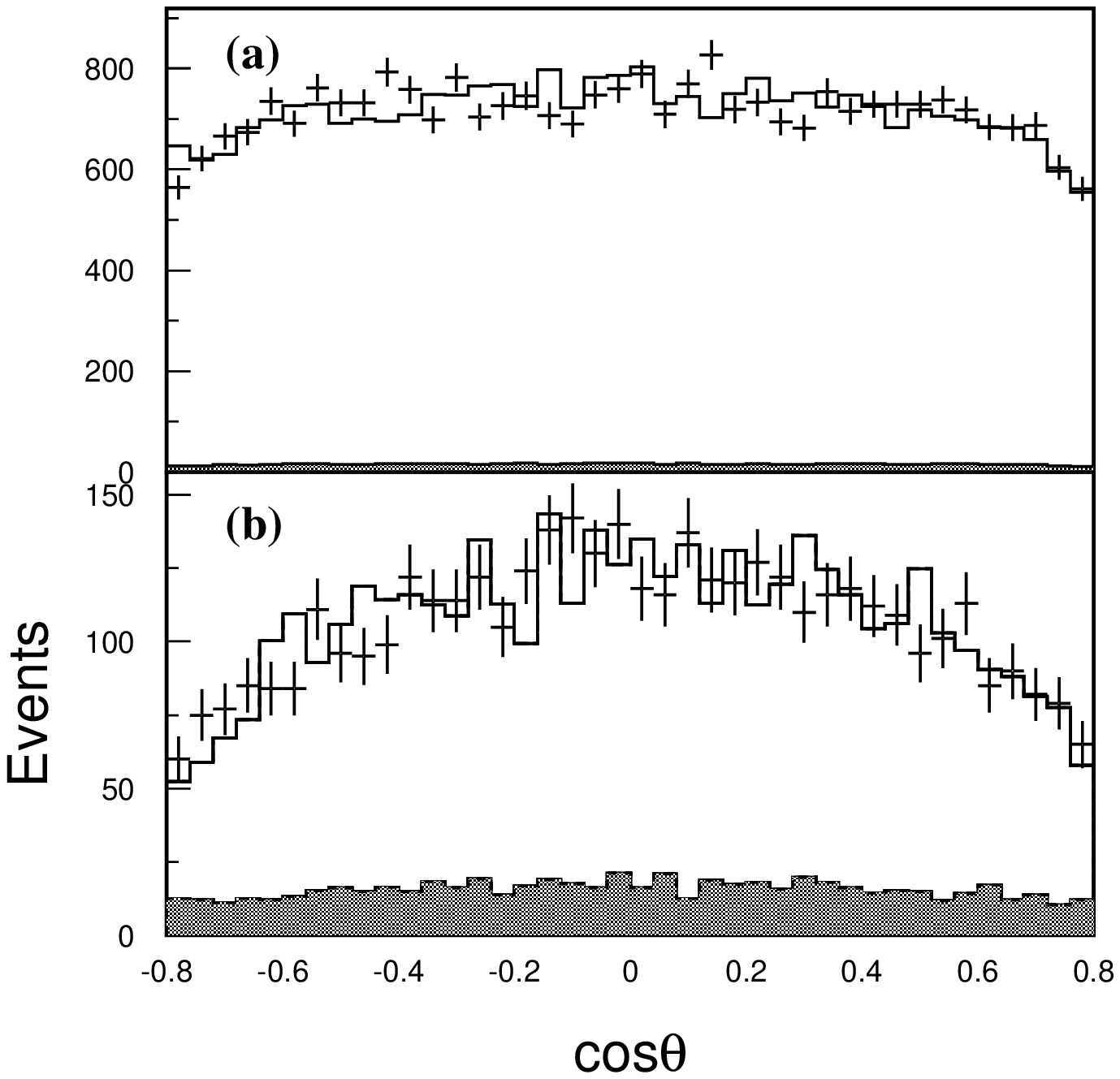} 
\caption{The $\cos\theta$ distribution of (a) charged tracks and (b)
  $\eta$ particles, where the crosses are data, the shaded histogram is from Monte
  Carlo simulation of $\jpsi\ar 3(\pip\pin)\eta$, and the full
  histogram is the sum of sideband background and Monte Carlo simulation.}
\label{6picos}
\end{center}
\end{minipage}
\end{figure}

Initially events were generated according to uniform phase space. 
However, the $\cos\theta$ distribution of charged tracks in $\jpsi\ar
2(\pip\pin)\eta$ was inconsistent with that from Monte Carlo
simulation. Much better agreement is obtained when the angular
distribution is generated according to $1+\alpha\cos^2\theta$, where
$\alpha=0.65\pm 0.03$ is obtained by fitting the $\cos\theta$
distribution of charged tracks.  Fig.~\ref{4picos}(a) and
Fig.~\ref{4picos}(b) show the comparison of the $\cos\theta$ distributions for charged
tracks and $\eta$, respectively, with Monte Carlo simulated events
with the charged tracks generated according to this distribution. 
Including the contribution
from $\eta$ sidebands, the angular distributions are
consistent.

For $\jpsi\ar 3(\pip\pin)\eta$, the 
detection efficiency is obtained from phase 
space events since  the $\cos\theta$ distributions
of charged tracks and $\eta$ are consistent with those from Monte
Carlo simulation,  as shown in Fig.~\ref{6picos}(a) and
Fig.~\ref{6picos}(b), respectively.
For the above two decay
modes, the detection efficiencies are listed in Table~\ref{result}.

\section{Systematic errors }
\label{syserreta}
The systematic errors mainly come from the following sources:

~~(1) The MDC tracking efficiency has been measured using channels like
$\jpsi\ar\Lambda \bar{\Lambda}$ and
$\psi(2S)\ar\pip\pin\jpsi, \jpsi\ar\mu^+\mu^-$. It is found that the Monte
Carlo simulation agrees with data within 1-2\% for each charged track.
Therefore,
8\% and 12\% are taken as the systematic errors in the tracking
efficiencies for the
4-prong and 6-prong final states analyzed here.


~~(2) The photon detection efficiency has been studied with different 
methods using $\jpsi\ar\rhoo\pio$ events~\cite{pheff}; the difference
between data and Monte Carlo simulation is about 2\% for each photon. In
this analysis, 4\% is taken as the systematic error for the $\eta$
decaying into  two photons.

~~(3) The kinematic fit is useful to reduce background.  The
systematic error from the kinematic fit is studied using the clean
channel $\jpsi\ar\pip\pin\pio$, as described in Ref.~\cite{rhopi}. The
efficiency difference of the kinematic fit between data and Monte
Carlo simulation is about 4\%. With the same method, the decay modes
$\jpsi\ar 2(\pip\pin)\pio$ and $\jpsi\ar 3(\pip\pin)\pio$ are also
analyzed. The efficiency difference of kinematic fit between data and
Monte Carlo are 4.3\% and 5.5\% respectively. Since $\jpsi\ar
2(\pip\pin)\pio$ and $\jpsi\ar 3(\pip\pin)\pio$ are similar to the two
channels analyzed in this paper, 4.3\% and 5.5\% are taken as the
systematic error of the kinematic fit.

~~(4) Other possible $\jpsi$ decay modes which may contribute to $\eta$ signals have
been studied, and the background from them can be ignored. The error
from the background under the $\eta$ peak is included in the fitting
error.  The uncertainties of the background shape in the two
channels are estimated to be about 3.4\% by changing the order
of the polynomial. Possible background
from the continuum events~\cite{yuancz}
is estimated using data at $\sqrt{s}=3.07$ GeV/c. 
After applying the same selection criteria as above, no significant 
$\eta$ signal is observed. Therefore, the background from this source
is also negligible. From the above analysis, the background uncertainty
for
the two decay modes is less than 5\%, which is taken as the background
systematic error.


~~(5) The branching fraction of $\eta\ar\g\g$ is $(39.43\pm0.26)\%$ \cite{pdg2004}. The error is also taken as a
systematic error.


~~(6) The number of $\jpsi$ events is $(57.70\pm 2.72)\times 10^6$, determined  from
inclusive 4-prong hadrons~\cite{fangss}. The uncertainty,
4.7\%, is
also a systematic error. Table \ref{syserr}
lists
the systematic errors from all sources.

\begin{table}[h] \centering
\caption{\label{toterr}Summary of systematic errors (\%)}

\begin{tabular}{l|c|c}
\hline
\hline
Source &   $2(\pip\pin)\eta~$ & $3(\pip\pin)\eta~$   \\\hline

MDC Tracking &   8 &12  \\

Photon effieciency &4 & 4 \\
Kinematic fit &4.3 & 5.5  \\


Background & 5 &  5 \\

$B(\eta\ar\g\g)$ & 0.7 & 0.7\\
Number of  $\jpsi$ evetns  & 4.7&4.7\\
\hline
Total  & 12.1 &15.4 \\
 \hline
\hline
\end{tabular}
\label{syserr}\end{table}


\section{Results}

The branching fractions are calculated with the following relation:
\begin{equation}
B(\jpsi\ar n(\pip\pin)\eta)={\frac{N_{obs}}{\eff\cdot B(\eta\ar\g\g)\cdot
N_{\jpsi}}},
\end{equation}

where n is 2 or 3, $N_{obs}$ is the observed events, $\eff$ is the detection
efficiency, $B(\eta\ar\g\g)$ is the branching fraction of  $\eta\ar\g\g$,
and $N_{\jpsi}$ the total number of $\jpsi$ events.

Table~\ref{result} 
summarizes the quantities used in the calculation  of the two
branching fractions and the final results, including systematic errors.
\begin{table}[htpb]
\caption{Numbers used and branching fractions measured.}
\begin{center}
\begin{tabular}{c|c|c|c}
\hline
Decay Modes   & $N_{obs}$ & $\eff$(\%)  & Branching Fraction \\
\hline
$\jpsi\ar 2(\pip\pin)\eta$ & 4839$\pm$158  & 9.43$\pm$0.10  &
$(2.26\pm 0.08 \pm 0.27)\times 10^{-3}$   \\

$\jpsi\ar 3(\pip\pin)\eta$ & 616$\pm$82  & 3.74$\pm$0.06    &
$(7.24\pm 0.96 \pm 1.11)\times 10^{-4}$   \\
\hline
\end{tabular}
\end{center}
\label{result}
\end{table}

\section{Summary}

In this paper, the decays of $\jpsi\ar 2(\pip\pin)\eta$ and $\jpsi\ar
2(\pip\pin)\eta$ are studied with the BESII $5.8\times 10^7$ $J/\psi$
event sample and their branching fractions are 
measured for the first time to be:

\begin{center}
$B(\jpsi\ar 2(\pip\pin)\eta)= (2.26\pm0.08\pm0.27)\times 10^{-3}$
\end{center}
\begin{center}
$B(\jpsi\ar 3(\pip\pin)\eta)= (7.24\pm0.96\pm1.11)\times 10^{-4}$
\end{center}

Comparing with other branching fractions of $\jpsi$ decaying into stable
hadrons, the branching fractions of $\jpsi\ar 2(\pip\pin)\eta$ and 
$\jpsi\ar 3(\pip\pin)\eta$ are not large. 



   The BES collaboration thanks the staff of BEPC and the computing 
center for their hard efforts.
This work is supported in part by the National Natural Science Foundation
of China under contracts Nos. 19991480, 10225524, 10225525, the Chinese Academy
of Sciences under contract No. KJ 95T-03, the 100 Talents Program of CAS
under Contract Nos. U-11, U-24, U-25, and the Knowledge Innovation Project
of CAS under Contract Nos. U-602, U-34 (IHEP); and by the
National Natural Science Foundation of China under Contract
No.10175060 (USTC), and No. 10225522 (Tsinghua University); and by the
U. S. Department of Energy under Contract N0. DE-FG02-04ER41291.


\begin{thebibliography}{120}
\bibitem{aube} J. J. Aubert,  et al., Phys. Rev. Lett. 33 (1974) 1404.
\bibitem{augu} J. E. Augustin, et al., Phys. Rev. Lett. 33 (1974) 1406.
\bibitem{physrep} L. Kopke and A. Wermes, Phys. Rep. 174 (1989) 67.
\bibitem{besii} J. Z. Bai, et al., Nucl. Instrum. Methods,
 A458 (2001) 627.
\bibitem{rhopi} J. Z. Bai, et al., Phys. Rev. D70 (2004) 012005.

\bibitem{pdg2004}S. Eidenlman, et al.,(Particle
Data Group), Phys. Lett. B592 (2004) 1, and references therein.

\bibitem{pheff} S. M. Li, et al., High Energy Phys. and Nucl. Phys. 28
(2004) 859 (in Chinese).
\bibitem{yuancz} P. Wang, X. H. Mo, C. Z. Yuan, Phys. Lett.  B557 (2003)
192 

\bibitem{fangss} S. S. Fang,  et al., High Energy Phys. and Nucl. Phys. 27,(2003)
277 (in Chinese).


\end{thebibliography}
\end{document}